\documentclass{aa}
\usepackage{graphicx}

\begin{document}

\title{Eclipse observations of high-frequency oscillations in active 
       region coronal loops}
      
\author{A.C.\ Katsiyannis\inst{1}, D.R.\ Williams\inst{2,1}, 
        R.T.J.\ M$^{\mathrm{c}}\!$Ateer\inst{1}, P.T.\ Gallagher\inst{3,1}, 
        F.P.\ Keenan\inst{1}, F.\ Murtagh\inst{4,5}}

\offprints{A.C. Katsiyannis, \email{A.Katsiyannis@qub.ac.uk}}

\institute{Department of Pure and Applied Physics, 
           Queen's University Belfast, Belfast, BT7 1NN, U.K.  \and
           Mullard Space Science Laboratory, University College
           London, Holmbury St. Mary, Dorking, Surrey, RH5 6NT, UK
           \and L-3 Communications EER Systems Inc., NASA Goddard
           Space Flight Center, Greenbelt, MD 20771, U.S.A. \and
           School of Computer Science, Queen's University Belfast,
           Belfast BT7 1NN, UK \and Observatoire Astronomique de
           Strasbourg, 11, rue de l'Universit\'e 67000 Strasbourg,
           France.}

\date{}

\abstract{
  One of the mechanisms proposed for heating the corona above solar
  active regions is the damping of magnetohydrodynamic (MHD)
  waves. Continuing on previous work, we provide observational
  evidence for the existence of high-frequency MHD waves in coronal
  loops observed during the August 1999 total solar eclipse. A wavelet
  analysis is used to identify twenty $4\times4$ arcsec$^2$ areas
  showing intensity oscillations. All detections lie in the frequency
  range 0.2--0.3~Hz (5--3~s), last for at least 3 periods at a
  confidence level of more than 99\% and arise just outside known
  coronal loops. This leads us to suggest that they occur in low
  emission-measure or different temperature loops associated with the
  active region.

\keywords{MHD -- waves -- eclipses -- Sun: activity -- Sun: corona -- 
          Sun: oscillations}

}

\authorrunning{A.\ C.\ Katsiyannis et al.}
\titlerunning{High-frequency oscillations in coronal loops}
\maketitle

\section{Introduction}

The coronal heating mechanism is the subject of a great deal of
debate. With a temperature of more than a million degrees, the corona
is several orders of magnitude hotter than the photosphere and
chromosphere, thus ruling out the possibility of heating via thermal
conduction. Popular theories which attempt to explain coronal heating
can be broadly grouped into two categories (see the review article by
Priest \& Schrijver 1999). One possibility is that a large number of
magnetic reconnections, followed by current dissipation, result in
frequent micro- or nano-flare activity (Parker 1988). The other theory
argues that the heating is dominated by the damping of
magnetohydrodynamic (MHD) waves -- either propagating from the lower
solar atmosphere or induced in active regions by reconnection --
through ion viscosity and electrical resistivity (first introduced by
Hollweg 1981). MHD waves have two very distinct extremes:
magnetoacoustic (divided into slow and fast mode) and `pure' Alfv\'en
(divided into compressional and non-compressional)
waves. Magnetoacoustic waves cause pressure variations in the coronal
plasma as they propagate. By contrast, Alfv\'en waves will either be
transverse and non-compressional, propagating parallel to the magnetic
field, or compressional and modifying the magnetic flux density
perpendicular to the field. The differences between the two main
categories of waves are expected to be observable, since Alfv\'en
waves cause only Doppler shifts in observed lines, whereas the
magnetoacoustic waves are expected to cause intensity variations as
well. The latter should be more readily observable since the intensity
normally varies with the square of the electron density.

In the past, a number of authors have reported intensity, velocity and
line width fluctuations in the corona. Koutchmy et al.\ (1983) used
coronagraph observations to find evidence of fluctuation in the
velocity measurements of the coronal line of \ion{Fe}{xiv} at 5303
{\AA} with periods of 300s, 80s and 43s. Their observations were
limited by sky fluctuations, which led to their suggestion of using
satellite telescopes or observations during solar total eclipse.
Furthermore Pasachoff \& Landman (1984), using observations made
during a total eclipse, detected intensity fluctuations with
frequencies in the range 0.5--2.0~Hz (periods of 0.5--2.0~s). Since
then, more detections of possible MHD oscillations have been published
and Aschwanden et al. (1999) produced a catalogue of all detected
periodicities across the spectrum from 0.01 to 1000~s. Subsequently, a
number of authors have reported further detections of coronal
oscillations. Singh et al.\ (1997) observed fast magnetosonic
disturbances with frequencies of 0.2 Hz during the 1995 total solar
eclipse. Cowsiket et al.\ (1999) applied similar techniques to detect
oscillations with frequencies in the range of 10-20 mHz during the
1998 total solar eclipse, while Sakurai et al.\ (2002) used
spectroscopic data to detect Doppler velocity with a frequency of
$\sim$57 mHz. More recently, total eclipse observations were published
by Pasachoff et al.\ (2002), who found some evidence for waves with
frequencies in the range of 0.75-1.0 Hz.

Porter et al.\ (1994a, b) have used numerical methods to simulate the
damping of energy from slow- and fast-mode MHD waves. They concluded
that slow-mode waves can deposit enough energy to heat the corona
under certain conditions, and for periodicities $\tau
\leq$100~s; for fast-mode waves the upper limit to periodicities is
$\sim$1~s. In a bid to detect such high-frequency oscillations,
Phillips et al.\ (2000, hereafter P00) developed the Solar Eclipse
Coronal Imaging System (SECIS), an instrument capable of making
high-cadence observations of solar eclipses. Williams et al.\ (2001,
2002, hereafter W01 and W02 respectively) reported SECIS detections of
oscillations with frequencies around 0.1 Hz, which might provide
enough energy to heat the corona efficiently. Here we report further
instances of such high-frequency oscillations in the data as those
presented by W01 and W02.

\section{Data analysis \& Results}

\subsection{Observations}

\begin{figure}
\centering
\includegraphics[bb=50 150 460 580, width=8.5cm]{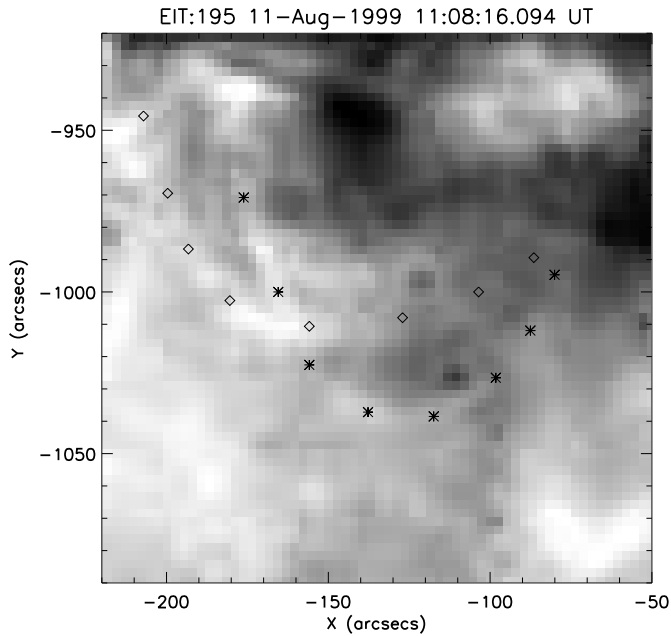}
\caption{Marked with diamonds and asterisks are the two coronal 
         loops in active region NOAA AR 8651 as observed by EIT in
         Fe~{\sc xii} (195 {\AA}) formed at around 1.5 MK. The image
         is rotated clockwise by 123 deg to coincide with the
         orientation of the SECIS data. Both loops were also observed
         by SECIS and analysed for coronal oscillations.}
\label{eit_image}
\end{figure}

Observations of the solar corona in Fe~{\sc xiv} 5303 {\AA} (formed at
$\sim$2.0 $\times$10$^{6}$~K during the August 1999 eclipse were taken
using the SECIS instrument (for a detailed description of the
instrument see P00 and W01), with a sampling rate of 44 frames per
second. The field of view of the instrument was $0.5
\times 0.5$ deg${^2}$ with a pixel size 4.0 arcsec pixel$^{-1}$. To
increase the signal-to-noise, for each pixel the intensities of the
eight adjacent pixels were added to it; i.e. as in W01 and W02 we have
summed over a 3$\times$3 pixel$^{2}$ area. Two active region loops in
NOAA AR 8651, observed both by the Solar and Heliospheric Observatory
(SoHO) and through an \ion{Fe}{xiv} (5303 {\AA}) filter by SECIS, are
highlighted in Fig.\ 1. Both loops were chosen as they are among the
most well isolated in this active region and can clearly be seen in
data from both SECIS and the EUV Imaging Telescope (EIT) on board
SoHO. The latter observations were used to confirm the positions of
these loops using several reference features.

\subsection{Wavelet analysis}

As in W01 and W02, we chose to analyse the data using a continuous
wavelet analysis (for further details on this technique see Torrence
\& Compo 1998, hereafter TC98). Although Fourier analysis is overall
more widely used, wavelet analysis has recently gained popularity, due
to its ability to detect oscillations localised both in time and
frequency. If an oscillation only lasts for a small fraction of the
time series duration, Fourier analysis will be unable to detect it,
whereas wavelet analysis is sensitive to even transient oscillatory
signals. As the oscillations we hope to detect are relatively short
(of the order of a few seconds) and our data extend over a period of
around 40 s, wavelet analysis is ideal. Other authors (e.g. Gallagher
et al.\ 1999; Ireland et al.\ 1999; Banerjee et al.\ 2000) have
applied the same technique to detecting solar coronal
oscillations. Additionally, the wavelet technique has consistently
been used by W01 and W02, and a comparison with the results from these
publications may help to draw interesting conclusions.

A Morlet wavelet was used for the analysis of our data, with

\begin{equation}
\psi(\eta)=\pi^{-1/4}\exp(i\omega_0\eta)\exp(\frac{-\eta^{2}}{2}),
\label{morlet}
\end{equation}

\noindent where $\eta = t/s$ is the dimensionless time parameter, $t$ 
is the time, $s$ the scale of the wavelet (i.e. its duration),
$\omega_0 = s\omega$ is the dimensionless frequency parameter, and
$\pi^{-1/4}$ is a normalization term (see TC98).

\subsection{Detections}

\begin{figure}
\centering
\includegraphics[bb=60 220 480 530, width=8.5cm]{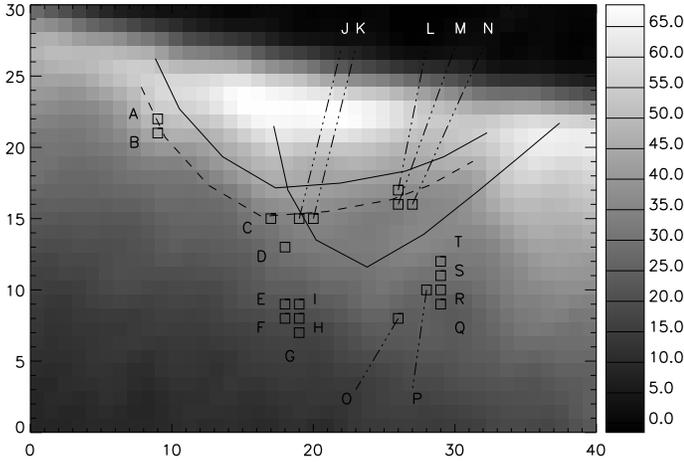}
\caption{NOAA AR 8651 loops as observed by SECIS. The gray scale 
        on the right-hand side corresponds to pixel counts. Only the
        tops of the loops shown in Fig.\ 1 are visible. The two solid
        lines highlight the two loops, while the squares contain
        pixels that show oscillations lasting three periods or
        longer. The 20 regions showing oscillations are labelled A to
        T. The dashed line is the solid line shifted by one pixel to
        the left and two pixels to the bottom, illustrating the good
        match of the shift to the pixels showing intensity
        oscillations.}
\label{secis_image}
\end{figure}

Fig.\ 2 contains a time integrated (40 s) SECIS image of approximately
the same region as Fig.\ 1. Since the lower parts of the corona are
covered by the Moon, only the apexes of the loops identified in EIT
Fe~{\sc xii} (195 {\AA}) are observed by SECIS. The solid lines
indicate the positions of the loops marked in Fig.\ 1 and the pixels
marked with squares the areas that show intensity oscillation lasting
three periods or longer. All twenty regions containing detected
oscillations are labelled from A to T. Importantly, all the detections
were made outside both bright loops toward the side where the corona
is more tenuous. Although the areas of the corona that appear to host
intensity oscillations are outside the visible loops, this has not led
us to believe that these waves perturb outside coronal loops. It could
simply be that the loops they travel through are low emission-measure
structures that are too faint to be visible above the
background. Furthermore, one may observe that in the case of the
left-hand loop, the detections exactly coincide with a shift of the
line which highlights the loop by one pixel to the left and two pixels
down. The dashed line of Fig.\ 2 illustrates the close correlation
between the shifted line of the loop and the pixels that have been
detected with long intensity oscillations. Also, phase analysis of
points A--T reveals no correlation between the phase of the
oscillations detected in each point in space or time. This leads us to
believe that this is not a simple case of a standing or travelling
wave.

\begin{figure}
\centering
\includegraphics[bb=20 20 575 739, width=8.5cm]{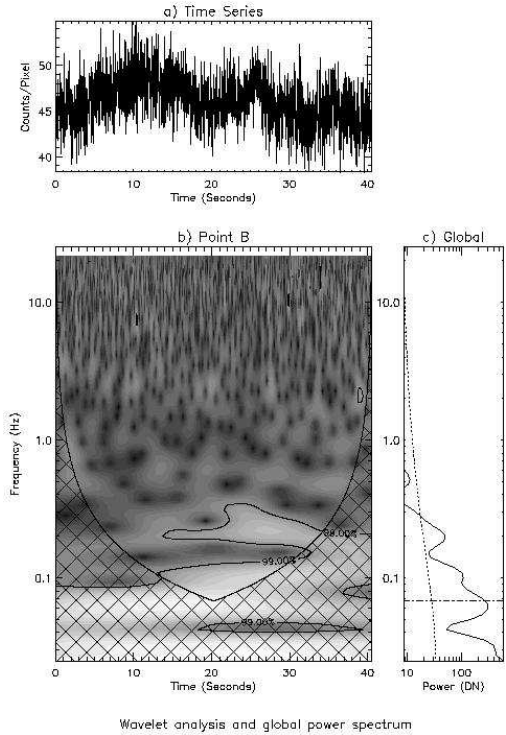}
\caption{Wavelet transform analysis for point B in Fig.\ 2. (a) is the 
         time series of the Fe~{\sc xiv} (5303 {\AA}) line
         observations, (b) contains the wavelet transform of the time
         series and (c) the global wavelet spectrum. The brighter an
         area in (b), the greater the oscillatory power at the given
         time and frequency. The contours in this panel highlight the
         areas where the detected power is at the 99\% confidence
         level. The hatched area of (b) represents the
         cone-of-influence (COI) and any detected oscillations within
         this region should be discarded as they might be influenced
         by edge effects. The scale of the frequency axis is
         logarithmic, while the time axis is linear and coincides with
         the time axis of (a). The dot-dashed line in (c) is a mark of
         the lowest limit of the COI and the dashed line the 99\%
         significance level (as the contours of panel (b)). Both the
         power and frequency axes of (c) are logarithmic.}
\label{Point B}
\end{figure}

\begin{figure}
\centering
\includegraphics[bb=20 20 575 739, width=8.5cm]{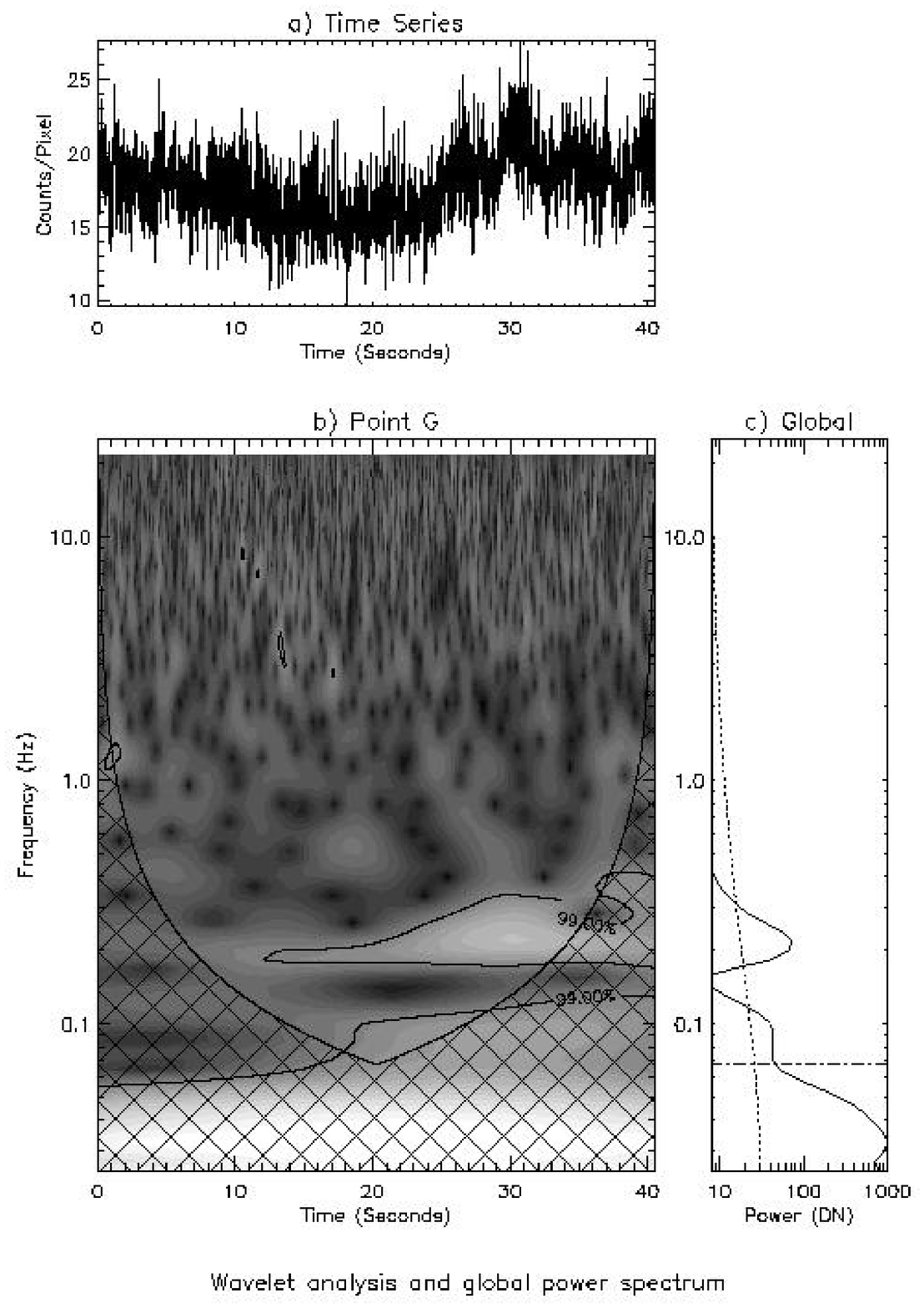}
\caption{The wavelet analysis of point G in Fig.\ 2. For more details 
                  see caption of Fig.\ 3}
\label{Point G}
\end{figure}

\begin{figure}
\centering
\includegraphics[bb=20 20 575 739, width=8.5cm]{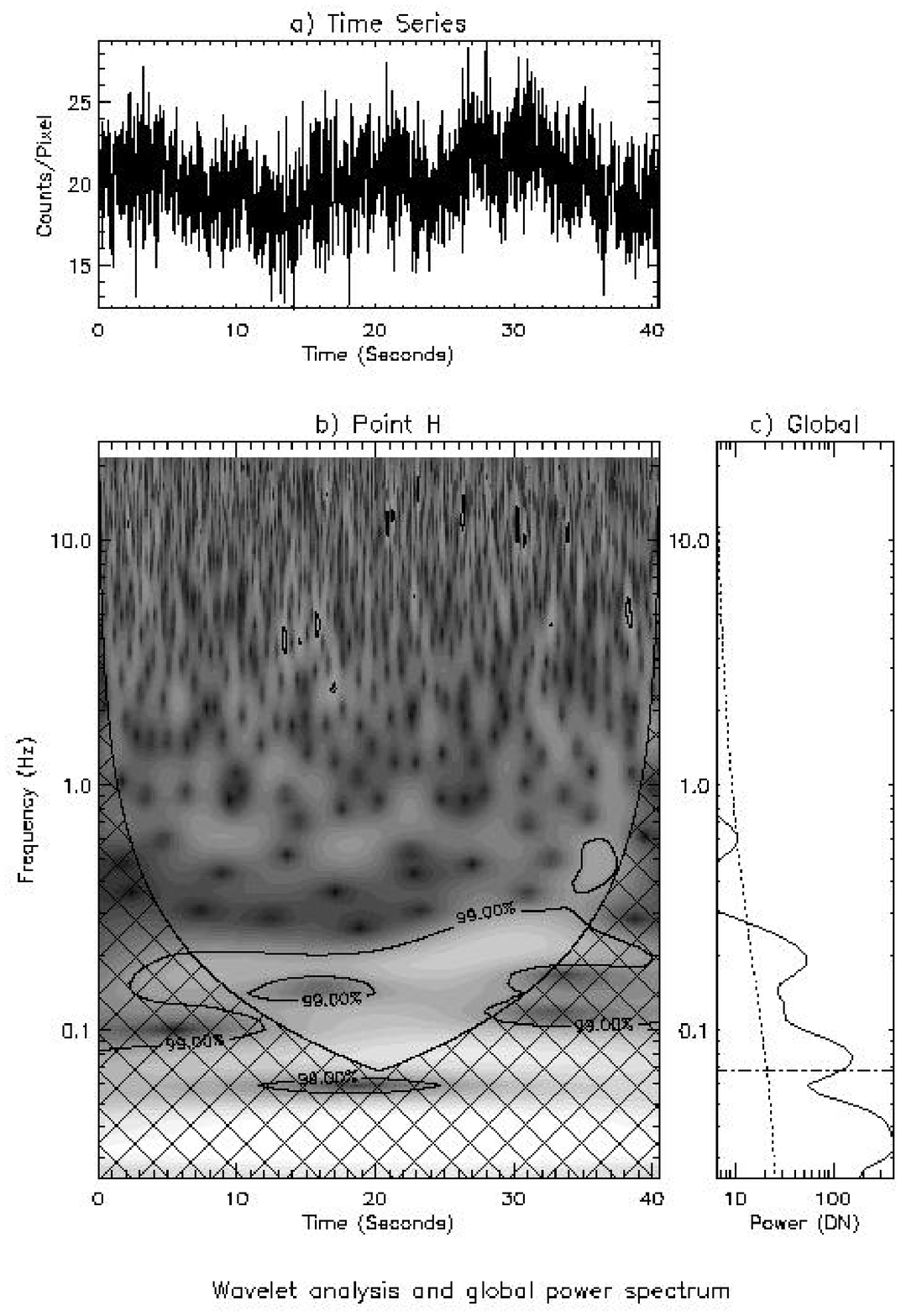}
\caption{The wavelet analysis of point H in Fig.\ 2. For more details 
                  see caption of Fig.\ 3}
\label{Point H}
\end{figure}

\begin{figure}
\centering
\includegraphics[bb=20 20 575 739, width=8.5cm]{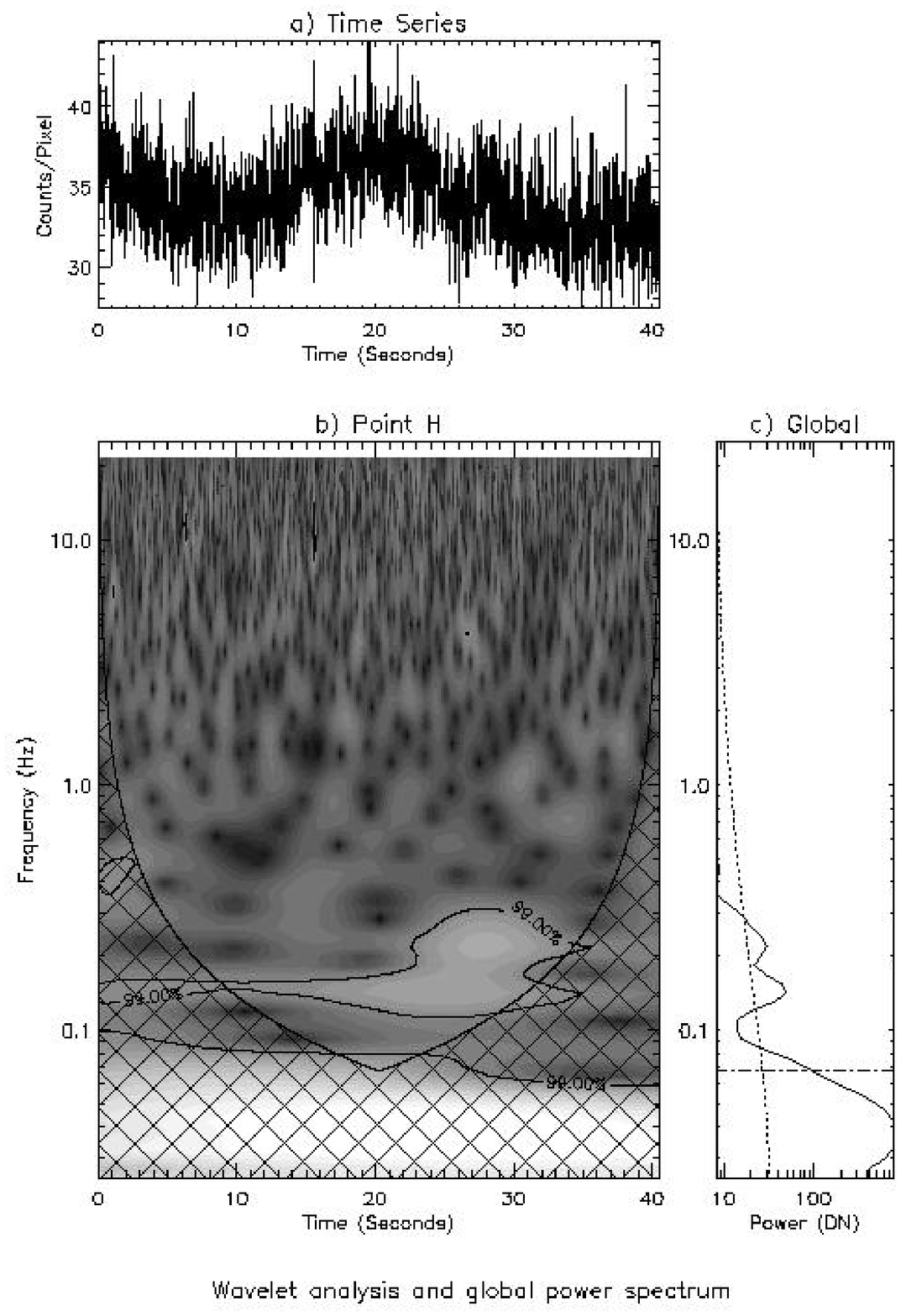}
\caption{The wavelet analysis of point N in Fig.\ 2. For more details 
                  see caption of Fig.\ 3}
\label{Point N}
\end{figure}

Figs.\ 3--6 contain the results of the wavelet transform of a sample
of points A to T in Fig.\ 2. The remaining figures are available from
the first author upon request. Each figure is divided into three
sections. Part (a) shows the time series generated by the Fe~{\sc xiv}
(5303 {\AA}) line filter (for details about the filter and the data
reduction see W01). Part (b) is the power density wavelet transform of
the time series in (a) : the brighter an area, the greater the power
at the given time and frequency. The vertical axis is logarithmically
scaled in frequency space while the horizontal axis is time on a
linear scale. The time axis in (b) exactly coincides with that of the
time series in (a), with the hatched region of (b) marking the
cone-of-influence (COI). Everything inside the COI should be treated
with suspicion, since any detections in this area may be influenced by
edge effects in the wavelet transform calculation. For example a
detection of a 0.5~Hz oscillation 0.5 s before the end of the time
series is unreliable as there is not sufficient time for the
oscillation in the wavelet packet to finish. For a more detailed
discussion of the problem see TC98. The contours of panel (b) indicate
where detected power exceeds the 99\% confidence level, i.e., there is
a 1\% chance of the detection being due to Poisson noise.

Panel (c) contains the global wavelet spectrum, which is the wavelet
analogue of the standard Fourier transform. It is produced by summing
the power density wavelet transform over the whole time series, while
the dotted line running along the frequency axis is the global
significance level (again summed over time) at the same value (99\%)
as the contours in panel (b). The horizontal dot-dashed line near the
bottom of panel (c) marks the bottom of the COI and all detections
below this frequency should be discarded. Since the third panel
provides no time information for the detections, the rest of the COI
cannot be defined.

At this point it should be emphasized that there are two additional
basic differences between panel (c) and the `traditional' Fourier
transform. Firstly, the spectrum is much smoother than the discrete
Fourier power spectrum. Secondly, due to the wavelet transform's use
of `Heisenberg boxes' (Mallat 1998), the better defined the transform
is in time, the less well defined it will be in frequency.

Table 1 contains all the frequencies detected in each of the points
from A to T. The length of these oscillations is also included in
units of periodicities (i.e., a duration of three means that this
particular oscillation lasted for three oscillatory periods at this
frequency).

\begin{table*}
\centering

\begin{tabular}{||c||c|c|c|c|c|c|c|c|c|c|c|c|c|c|c|c|c|c|c|c||}
\hline 
Point & A & B & C & D & E & F & G & H & I & J & K & L & M & N & O & P & Q & R & S & T\\
\hline 
Period (s) & 5 & 5 & 5.5 & 5 & 4  & 5 & 4.5 & 5 & 5.5 & 4.5 & 5 & 6.5 & 6.5 & 7 & 6  & 5 & 5 & 5 & 6 & 4 \\
\hline 
Duration (no. of periods) & 4 & 4 & 3 & 3 & 4 & 3 & 4 & 5 & 3 & 3 & 3 & 3.5 & 3.5 & 3 & 3 & 3 & 3 & 3 & 3 & 3\\
\hline
\end{tabular}

\caption{Frequencies detected in each of the points A 
                  through T. The duration of these oscillations is
                  also included in units of Morlet periodicities
                  (i.e. a duration of three means that this particular
                  oscillation lasted for three oscillatory periods at
                  this frequency).}
\label{table1}
\end{table*}

\section{Discussion}

Although W01 and W02 have published similar detections from SECIS
observations, the present work is by far the largest number of
detections published to date. All twenty points presented here (Table
1) have passed a number of selection criteria. These include those
used by W01 and W02 and the most important are summarised below:

\begin{itemize}

\item The frequencies of the detections are distinct from known 
      instrumental frequencies (see W01).

\item The contours of panel (b) were chosen at a 99\% confidence 
      level. Only oscillations within those contours were considered.

\item All the reported detections lasted for at least three 
          periods, so as to rule out rapid increases or decreases in
          the signal. When the duration of the oscillations was
          calculated, any portion within the COI was discarded.

\end{itemize}

As with all high-cadence systems producing data analysed for
oscillations, the introduction of instrumental frequencies and the
effect of noise is always a concern. To address the first limitation,
wavelet analysis was deployed for parts of the image covered by the
moon's disk and the very faint parts of the corona. Any non-localised
instrumental variations in pixel intensity should have affected these
areas as much as the pixels of the active region. With the exception
of those instrumental frequencies that are already known and discussed
by W01, no other frequencies were detected. The other well-know cause
of false detections is noise. To limit the possibility of a false
detection because of noise, we chose to ignore any detections with
frequencies above 1 Hz. It is widely accepted (for example Starck \&
Murtagh (2002) and references therein), that Gaussian or Poisson noise
only affects frequencies of the same order as the sampling rate of the
time series. As the sampling rate of the SECIS 1999 observations was
44 frames per second, the reported frequencies should be fairly
unaffected by Gaussian or Poisson noise (which includes types of noise
such as the CCD readout).

The two coronal loops of AR 8651 analysed here show a significant
number of oscillations. Although only twenty oscillations were
detected lasting three periods or longer, several tens of wave
signatures were found which last between two and three periods at a
$>99$\% confidence level. This number is much larger than that
expected considering the previous work on SECIS data (W01 and W02).
Moreover, all previous detections were from the interior of bright
coronal loops, while all of our oscillations are detected in fainter
loops within the same active region, toward the tenuous part of the
corona. Both of the above peculiarities can be explained by
introducing a different physical mechanism to that suggested in
W02. Zaqarashvili \& Roberts (2002, 2003) have suggested a swing
wave-wave interaction mechanism which may cause energy transformation
from fast magnetoacoustic waves propagating across a magnetic field to
Alfv\'en waves propagating along the field.  They argue that, for a
given medium density and magnetic field, the energy of fast waves can
be converted to Alfv\'en waves with a basic harmonic at half the
wavelength of the fast-mode wave.  The above mechanism (also called
{\it swing absorption}) may provide a possible explanation for the
intensity oscillations reported here.  In this scenario, Alfv\'en
waves created in the upper photosphere (as described by Zaqarashvili
\& Roberts 2002, 2003) propagate along magnetic field lines adjacent
to the coronal loops in Figs.\ 1 and 2. In the case of the first loop
(points A, B, C, J, K, L, M, N of Fig.\ 2), a magnetic field line was
running almost parallel to the left bright loop of Fig.\ 1,
producing the alignment highlighted in Fig.\ 2. The rest of the points
with detected oscillations (D--I and O--T) belong to one or another of
the magnetic field lines in the same active region. 

Litwin \& Rosner (1993) proposed a multi-thread model, where many tiny
loops with different physical parameters but in steady state
equilibrium are superimposed forming the observed coronal loop
structures.  Aschwanden et al.\ (2000) used this model to explain the
nonuniform heating of coronal loops observed by the {\it Transition
Region And Coronal Explorer} (TRACE). Several such small threads may
form thinner and fainter loops outside the bright structures appearing
on Fig.\ 1 and 2. As the above oscillations travel through low
emission-measure loops, it is easier for us to detect more
oscillations than through high emission-measure loops (such as those
studied by W01 and W02). Therefore the conditions under which those
MHD waves propagate along points A--T are significantly different from
those of the oscillations reported by W01 and W02. In their case the
emission from the propagating waves is much stronger and as they and
travel through a large number of threads, they stand significantly
above the background emission of the loop. This is more clearly seen
in W02, where the event that caused the oscillations also cause them
to propagate with the same phase, enabling W02 to calculate the
velocity of the perturbation by the phase difference of the traveling
wave across the loop. In contrast to those detections, the area
outlined by points A--T contains a smaller amount of threads with
lower emission-measure, therefore relatively weaker MHD oscillations
can be detected. This is supported by the results of the phase
analysis that reveal no correlation in the phase of the 20 points with
detected oscillations. The mechanism that produced the weaker waves
(compared to those reported by W01 and W02) is less likely to produce
them with the same phase. As the single-phase MHD waves are relatively
rare events in the solar corona, we would expect them to appear in
more extreme conditions (such as the release of large amounts of
energy at the foot points of the loops) while the propagation of
weaker oscillations through a smaller number of threads is more likely
to take place under phase mixing conditions.

Cooper et al.\ (2003) suggest a possible mechanism that explains in
some detail the detection of intensity oscillations as a line-of-sight
effect of entirely incompressible MHD waves. In this model, when
observed at an angle $\theta$ to the direction of propagation, the
wave-induced deformation in a coronal loop causes intensity
variations. This is because the amount of optically thin emitting
plasma along the line of sight changes as a function of time. Cooper
et al.\ (2003) find that the observed amplitude of the intensity
oscillation can vary as a function not only of the true intensity of
the oscillation and the angle between the propagation direction and
the line-of-sight, but also of the wavelength of the
perturbation. Furthermore, the observed frequency also varies as a
function of the angle $\theta$, meaning that the detected periods
listed in Table 1 may simply be higher harmonics of the true values.

Using the Cooper et al.\ (2003) and Zaqarashvili \& Roberts (2002)
results we were able to provide a satisfactory explanation of how the
detected incompressible MHD waves were created in the vicinity of an
active region in the photosphere, transmitted through low
emission-measure loops to the lower corona and then detected as
intensity oscillations by our imaging system.  The large number
(comparing to previous work) of detections and the alignment of some
of these can be explained as the low emission that comes from the
tenuous plasma makes any intensity oscillations more apparent, since
the oscillating material makes for a higher percentage of the detected
intensity.

\begin{acknowledgements} 

The authors would like to thank K.J.H. Phillips for his collaboration
on the SECIS project. ACK acknowledges funding by the Leverhume Trust
via grant F00203/A. DRW acknowledges a CAST studentship funded by the
Department of Employment \& Learning and the Rutherford Appleton
Laboratory. JMCA acknowledges CAST studentship funded by DEL and
QUB. ACK \& DRW would like to thank Valery Nakariakov and Temury
Zaqarashvili for useful discussions.

\end{acknowledgements}

\end{document}